\documentclass[11pt,letterpaper]{article}
\usepackage[margin=1in]{geometry}
\usepackage[T1]{fontenc}
\usepackage{lmodern}
\usepackage{amsmath,amssymb,amsthm,mathtools}
\usepackage{microtype}
\usepackage{booktabs,array,tabularx}
\usepackage{enumitem}
\usepackage{tikz}
\usetikzlibrary{arrows.meta,positioning,shapes.geometric}
\usepackage[numbers,sort&compress]{natbib}
\usepackage{xurl}
\usepackage[colorlinks=true,linkcolor=blue!45!black,citecolor=blue!45!black,urlcolor=blue!45!black]{hyperref}
\usepackage[nameinlink,noabbrev]{cleveref}
\hypersetup{pdftitle={On the Complexity of Minimal CNOT Circuit Synthesis},pdfauthor={Chaoyang Qin, Jialiang Tang, Xiaohan Shan, Xiaoming Sun}}
\setlist{itemsep=3pt,topsep=5pt}
\newtheorem{theorem}{Theorem}[section]
\newtheorem{lemma}[theorem]{Lemma}
\newtheorem{corollary}[theorem]{Corollary}

\theoremstyle{definition}

\newtheorem{example}[theorem]{Example}

\newcommand{\F}{\mathbb F_2}
\newcommand{\GL}{\operatorname{GL}}
\newcommand{\wt}{\operatorname{wt}}
\newcommand{\tw}{\operatorname{tw}}
\newcommand{\NP}{\mathsf{NP}}
\newcommand{\PP}{\mathsf{P}}
\newcommand{\cnot}[2]{#1\!\to\!#2}
\newcommand{\row}[2]{(#1)_{#2,*}}

\newcommand{\PinColor}{\mathsf{PinnedBipartite3EdgeColor}}
\newcommand{\indicator}{\mathbf 1}

\title{On the Complexity of Minimal CNOT Circuit Synthesis}
\author{%
Chaoyang Qin\textsuperscript{1,2}\quad
Jialiang Tang\textsuperscript{1,2}\quad
Xiaohan Shan\textsuperscript{3}\quad
Xiaoming Sun\textsuperscript{1,2,*}\\[6pt]
\small \textsuperscript{1}State Key Lab of Processors, Institute of Computing Technology,\\
\small Chinese Academy of Sciences, Beijing 100190, China\\[3pt]
\small \textsuperscript{2}School of Computer Science and Technology,\\
\small University of Chinese Academy of Sciences, Beijing 100049, China\\[3pt]
\small \textsuperscript{3}Beijing Zhongke Qhub Technology Co., Ltd.\ (Qhub), Beijing, China\\[3pt]
\small \textsuperscript{*}Corresponding author: \href{mailto:sunxiaoming@ict.ac.cn}{\texttt{sunxiaoming@ict.ac.cn}}
}
\date{}

\begin{document}
\maketitle
\vspace{-2em}
\begin{abstract}

Minimizing two-qubit gate depth in linear reversible circuit synthesis is critical for reducing latency in quantum compilation. We establish a sharp complexity threshold for exact CNOT synthesis under all-to-all connectivity and fixed wire labels. Given an explicit $n\times n$ binary matrix, we present a deterministic $O(n^2)$-time algorithm that decides whether it admits a circuit of depth at most two and constructs one whenever feasible; clean ancillas do not affect this feasibility. In contrast, deciding whether a target admits a circuit of depth at most three is $\NP$-complete. This hardness holds even for unit lower-triangular targets, both under any fixed polynomially bounded and polynomial-time computable clean-ancilla budget (including the ancilla-free case) and under unrestricted clean workspace. Our proof rests on a three-layer normal form for clean-ancilla circuits, which also shows that $\lfloor n/3\rfloor$ clean ancillas always suffice at depth three whenever any finite number does.

\end{abstract}

\section{Introduction}
\label{sec:introduction}

Quantum compilers routinely isolate blocks of CNOT gates and resynthesize them~\cite{GoubaultEtAl2021,MaslovZindorf2022,ShaikVanDePol2024}, because two-qubit gates are typically noisier and slower than single-qubit gates on current hardware. The parallel depth of such a block, i.e., the number of rounds of gates acting on disjoint qubits, directly determines its latency and its exposure to decoherence. During resynthesis, an optimizer may replace a block by \emph{any} shallower circuit implementing the same linear map. This raises a basic question: how hard is it to decide whether a given linear map admits a circuit of a prescribed small depth?

CNOT circuits compute invertible linear transformations over $\F$. Simulating such a circuit or checking the equivalence of two circuits is easy linear algebra, but synthesizing an optimal circuit for a prescribed transformation is much harder. Gate scheduling reorders a prescribed gate sequence, and peephole optimization rewrites short contiguous subcircuits. Exact synthesis, in contrast, receives only the target matrix $M\in\GL(n,2)$. A synthesis algorithm may introduce arbitrary intermediate parities, cancel earlier operations, and route information through auxiliary workspace. A hardness proof for exact synthesis must therefore rule out shortcuts through every possible resynthesis of the target map.

Minimizing parallel depth differs from minimizing gate count. A single layer can execute any set of CNOT gates whose supports are pairwise disjoint, whereas gates sharing a wire must be placed in different layers. Short sequential circuits can be deep, and gate-count reductions need not preserve layer structure, so the complexity of depth minimization does not follow from gate-count hardness. We study instance-specific depth minimization under all-to-all connectivity: given an explicit matrix $M\in\GL(n,2)$, can $M$ be synthesized within $k$ parallel CNOT layers?

Formally, for $M\in\GL(n,2)$ and a budget of $m\ge0$ clean ancillas, let $d_m(M)$ denote the minimum depth of a CNOT circuit $C$ on $n+m$ wires satisfying
\begin{equation}
 C(x,0^m)=(Mx,0^m)\qquad\text{for all }x\in\F^n.
 \label{eq:clean}
\end{equation}
We work in the standard fixed-label, all-to-all model without measurement. Any ordered pair of distinct wires may support a CNOT gate, wire permutations are not free, and the $m$ ancillas start at $0$, must return to $0$, and may be used for intermediate fanout and parity cancellation. Within these rules, a candidate circuit may resynthesize the target transformation from scratch.

\subsection{Our Results}

Our main contribution is a complexity classification of shallow CNOT synthesis with a sharp threshold between depths two and three.

\begin{theorem}[Tractability at Depth Two]
\label{thm:two}
Given an explicit binary matrix $M \in \F^{n \times n}$, one can decide in deterministic $O(n^2)$ time whether $M$ admits an ancilla-free CNOT implementation of depth at most two, and construct such a circuit whenever one exists. Moreover, clean ancillas provide no depth reduction at this threshold: for every $M\in\GL(n,2)$ and every $m\ge0$,
\[
 d_m(M)\le2\quad\Longleftrightarrow\quad d_0(M)\le2.
\]
\end{theorem}

The ancilla invariance in \cref{thm:two} follows directly from the cleanup constraint~\eqref{eq:clean} (\cref{lem:two-clean}), so the algorithmic content lies in the ancilla-free regime. The time bound refers to the word-RAM model with $O(\log n)$-bit words; reading the dense input matrix dominates the running time.

At depth three, exact synthesis becomes $\NP$-complete in every clean-ancilla regime.

\begin{theorem}[Hardness at Depth Three]
\label{thm:three}
For every fixed nonnegative integer-valued function $a(n)$ that is polynomially bounded and polynomial-time computable, deciding whether an explicit matrix $M \in \GL(n,2)$ satisfies $d_{a(n)}(M) \le 3$ is $\NP$-complete. Moreover, deciding whether there exists \emph{some} finite $m \ge 0$ such that $d_m(M) \le 3$ is also $\NP$-complete. In both settings, $\NP$-hardness holds even when the input $M$ is restricted to be unit lower-triangular.
\end{theorem}

A single reduction proves \cref{thm:three} for all ancilla budgets at once. Given an instance $G$ of a restricted bipartite edge-precoloring problem (whose precolored edges are called \emph{pins}), our construction produces an explicit matrix $M_G$, independent of the budget $m$, such that
\begin{equation}
 \forall m\ge0:\qquad
 d_m(M_G)\le3
 \quad\Longleftrightarrow\quad
 G\text{ admits a proper three-edge-coloring extending its pins}.
 \label{eq:uniform-reduction}
\end{equation}
Every yes-instance is therefore realizable in three layers without ancillas ($d_0(M_G) \le 3$), whereas no-instances cannot be synthesized in three layers even with an arbitrary finite supply of clean ancillas. This ancilla invariance is a feature of our reduction family. In general, clean ancillas do help at depth three: \cref{ex:ancilla} gives a five-wire matrix with $d_1(M)=3<d_0(M)=4$.

Our structural characterization of three-layer circuits also yields a uniform workspace bound, proved in \cref{sec:normal}.

\begin{theorem}[Workspace Saturation]
\label{thm:workspace}
For every $M\in\GL(n,2)$ and every $m\ge\lfloor n/3\rfloor$,
\begin{equation}
 d_m(M)\le3
 \quad\Longleftrightarrow\quad
 d_{\lfloor n/3\rfloor}(M)\le3.
 \label{eq:saturation-intro}
\end{equation}
\end{theorem}

\subsection{Related Work}
\label{sec:related}

\paragraph{Asymptotic and constructive synthesis.}
Moore and Nilsson showed that sufficient ancillary workspace enables logarithmic-depth parallelization of CNOT circuits~\cite{MooreNilsson2001}, and Patel, Markov, and Hayes established the asymptotically optimal worst-case gate count $\Theta(n^2/\log n)$ for ancilla-free synthesis~\cite{PatelMarkovHayes2008}. Jiang et al.~\cite{JiangEtAl2020} resolved the asymptotic space--depth trade-off: every $n$-qubit CNOT circuit can be synthesized with $m$ clean ancillas in depth
\[
 O\!\left(\max\left\{\log n,
 \frac{n^2}{(n+m)\log(n+m)}\right\}\right),
\]
matching counting lower bounds. They also posed the instance-specific depth minimization problem for a given matrix and ancilla budget; their hardness results concern topology-constrained synthesis and local peephole optimization, leaving open the complexity of exact depth minimization under all-to-all connectivity. On the constructive side, Goubault de Brugi\`ere et al.~\cite{GoubaultEtAl2021} developed divide-and-conquer and greedy depth-reduction methods, and Maslov and Zindorf~\cite{MaslovZindorf2022} gave constructive depth bounds for CZ, CNOT, and Clifford circuits. For restricted connectivity, linear-depth constructions and lower bounds are known for linear nearest-neighbor architectures~\cite{KutinMoultonSmithline2007} and for grids, ladders, and other hardware topologies~\cite{WuEtAl2023,GoubaultMartiel2025}.

\paragraph{Instance-specific gate count and structural bounds.}
Amy, Azimzadeh, and Mosca proved $\NP$-hardness for parity-network problems in which a prescribed collection of intermediate parities must appear during the computation~\cite{AmyAzimzadehMosca2018}, a requirement stricter than specifying only the final transformation; van de Wetering and Amy~\cite{VanDeWeteringAmy2024} established $\NP$-hardness of circuit optimization for several gate sets and cost metrics. In the all-to-all, fixed-label setting, concurrent preprints by Acuaviva, Acuaviva, and Acuaviva~\cite{AcuavivaEtAl2026} and by Li and Guan~\cite{LiGuan2026} prove that minimizing the CNOT \emph{gate count} is $\NP$-hard (the former also gives instances whose optimal gate count is invariant under clean or borrowed ancillas). Since gate count does not control layer packing, these reductions do not imply hardness for bounded-depth synthesis. Bu, Fan, and Joo~\cite{BuFanJoo2025} derive CNOT-count lower bounds, decide in $O(n^\omega)$ time whether a transformation with connected support graph can be synthesized with $n-1$ gates, and determine the exact depth of special families (e.g., cyclic permutations on at least seven wires have optimal depth six); J\o{}rgensen~\cite{Jorgensen2026} gives explicit $4n-o(n)$ gate-count lower bounds for concrete operators. Such results do not classify the complexity of deciding whether an arbitrary given matrix has a shallow implementation.

\paragraph{Exact and heuristic synthesis frameworks.}
Webster, Koutsioumpas, and Browne~\cite{WebsterEtAl2026} study heuristic and optimal synthesis of CNOT and Clifford circuits for gate count and depth. Shaik and van de Pol~\cite{ShaikVanDePol2024} encode exact CNOT synthesis for gate count and depth as SAT problems, with variants for fixed or permutable outputs and for layout restrictions; Peham et al.~\cite{PehamEtAl2023} give SAT-based depth-optimal Clifford synthesis, and Shaik and van de Pol~\cite{ShaikVanDePol2025} use Clifford normal forms for CNOT-count-optimal Clifford synthesis. Several of these frameworks allow terminal wire permutations or encoded ancillas~\cite{GoubaultEtAl2021,ShaikVanDePol2024,WebsterEtAl2026}, which differs from the fixed-label, restored-workspace model studied here. Exact solvers are effective on small instances but do not settle the polynomial-time complexity of the threshold between depths two and three.

\paragraph{Clifford depth hardness.}
Closest to our hardness result is a recent preprint by Keller and Kueng~\cite{KellerKueng2026}, who prove hardness of depth minimization over the Clifford gate set $\{X,Y,Z,H,S,S^\dagger,\allowbreak\mathrm{CNOT},\allowbreak\mathrm{CZ}\}$ by encoding cubic-graph edge coloring into phase-polynomial targets realized with native $\mathrm{CZ}$ gates. Hardness over a larger gate library does not imply hardness when both the target and the admissible gates are restricted to the $\F$-linear CNOT group. Our reduction encodes scheduling constraints purely into invertible linear maps over $\F$, and our soundness proof also accommodates arbitrary clean workspace through a three-layer normal form.

\subsection{Technical Overview}
\label{sec:overview}

\paragraph{Depth two: bounded treewidth from two matchings.}
Any two-layer ancilla-free circuit has the form $M = (I+Q)(I+P)$, where $P$ and $Q$ are the off-diagonal adjacency matrices of directed matchings. After identical gates appearing in both layers are canceled, every surviving gate leaves a visible entry in the off-diagonal support graph $H_M$ of $M$. If $J$ denotes the undirected union of the two reduced matchings, expanding the product gives
\[
 J\subseteq H_M\subseteq J^2.
\]
Since $J$ has maximum degree two, it is a disjoint union of paths and cycles, whose squares have treewidth at most four; hence $\tw(H_M)\le4$. Although $J$ is unknown, a width-four tree decomposition of $H_M$ can be computed in linear time. We then express two-layer synthesis as a constraint satisfaction problem on $H_M$: each wire receives one of at most $81$ local states encoding its roles in the two layers, and matching reciprocity and the row equations of $M$ become unary and binary constraints along edges of $H_M$. Dynamic programming over the tree decomposition decides feasibility and reconstructs a circuit.

\paragraph{Depth three: precolored edges against arbitrary resynthesis.}
If each edge of a bipartite graph corresponds to a single CNOT gate, a three-layer schedule is exactly a proper three-edge-coloring. The difficulty is that the synthesis problem receives only the product matrix $M$: a candidate circuit may combine sources into intermediate parities, route signals through unrelated wires, or use clean ancillas.

We reduce from $\PinColor$, a restricted bipartite edge-precoloring extension problem. In its instances $G=(S,T,E)$, every source in $S$ has degree three, every target in $T$ has degree one or three, and every precolored edge (a \emph{pin}) joins a source to a degree-one leaf while the source's other two neighbors have degree three. We derive its $\NP$-completeness from Fiala's theorem on edge-precoloring extension in cubic bipartite graphs~\cite{Fiala2003} using constant-size local gadgets (\cref{sec:pins}). Precoloring is indispensable: by K\H{o}nig's theorem~\cite{Konig1916}, every unpinned bipartite graph of maximum degree three is three-edge-colorable, and such a coloring can be found in linear time~\cite{ColeOstSchirra2001}.

Given $G$, we construct $M_G$ so that every source wire keeps its input, every main target accumulates the sum of its three neighboring sources, and each pin is replaced by a four-wire timing gadget, which we call a \emph{clock}. Soundness against arbitrary three-layer circuits with clean ancillas rests on four structural facts:
\begin{enumerate}[label=(\roman*)]
 \item \emph{Three-layer normal form:} every cleanup-compliant use of ancillas amounts to adding vertex-disjoint two-leaf outward stars to the middle data layer, with each star center idle in the two outer data layers.
 \item \emph{Source rigidity:} a wire whose output row is $e_i^\top$ and whose columns in $M$ and $M^{-1}$ have weight at least four is never a target in the normal form, so every source wire holds its input at every layer boundary.
 \item \emph{Main-target saturation:} every main target receives one direct injection from each of its three sources, in three distinct layers, and takes part in no other gate.
 \item \emph{Clock rigidity:} the column structure of each clock determines the layer in which its source injects, which enforces the pin color.
\end{enumerate}
None of these steps assumes that a candidate circuit respects the topology of $G$ or keeps gadgets separate; each follows from global algebraic invariants of $M_G$.

\paragraph{Organization.}
\Cref{sec:model} fixes the model and notation. \Cref{sec:two} gives the $O(n^2)$-time algorithm for depth two. \Cref{sec:normal} proves the three-layer clean-ancilla normal form, the workspace bound, and the source-rigidity lemma. \Cref{sec:reduction} presents the reduction from $\PinColor$ and its soundness against arbitrary resynthesis, and \cref{sec:complexity} derives $\NP$-completeness and its consequences for variable depth thresholds. \Cref{sec:discussion} discusses open problems.

\section{Preliminaries and Model}
\label{sec:model}

\paragraph{Linear algebra and circuit conventions.}
All vectors and matrices are over $\F$. We write $e_i$ for the $i$-th standard basis column vector, $E_{ji} = e_j e_i^\top$ for the matrix with a single $1$ at entry $(j,i)$, and $\row{M}{i}$ for the $i$-th row of $M$. The Hamming weight $\wt(v)$ of a vector $v$ is the number of its nonzero entries. Matrices act on column vectors from the left, so a circuit with layers $L_1, L_2, \dots, L_k$ (in chronological order) implements $L_k \cdots L_2 L_1$.

A CNOT gate $\cnot{i}{j}$ with control $i$ and target $j\ne i$ adds coordinate $i$ into coordinate $j$ and corresponds to $I + E_{ji}$. A \emph{matching layer} is a set of CNOT gates acting on pairwise disjoint sets of wires; its matrix is an involution. Empty layers (the identity) are allowed, so any circuit of depth at most $k$ can be padded to exactly $k$ layers.

\paragraph{Clean ancillas and problem formulations.}
The clean-ancilla condition~\eqref{eq:clean} constrains a circuit only on inputs whose $m$ ancilla coordinates are zero. In terms of the full $(n+m)\times(n+m)$ matrix, a clean-ancilla implementation of $M\in\GL(n,2)$ has the block upper-triangular form
\begin{equation}
 \widetilde M=\begin{pmatrix}M&B\\0&D\end{pmatrix},
 \qquad D\in\GL(m,2),
 \label{eq:extension}
\end{equation}
where $B\in\F^{n\times m}$ and $D$ are unconstrained. In particular, we do not require $\widetilde M=\operatorname{diag}(M,I_m)$ (cf.~\cite{AcuavivaEtAl2026}), so our lower bounds hold against the broadest class of clean-ancilla implementations.

We consider the following formulations of the decision problem:
\begin{itemize}
 \item In the \emph{fixed-budget, fixed-depth} regime (as in \cref{thm:three}), the budget function $a(n)$ and the depth threshold $k$ are fixed, and the input is the matrix $M\in\GL(n,2)$.
 \item In the \emph{variable-budget, variable-depth} regime, the input is $(M,1^m,k)$, with $m$ in unary (a polynomially bounded workspace) and $k$ in binary.
 \item In the \emph{unrestricted-ancilla} regime, the input is $M$, and we ask whether $d_m(M)\le k$ for some finite $m\ge0$.
\end{itemize}

\paragraph{Graph-theoretic definitions.}
All graphs used for matrix supports and edge colorings are simple and undirected; the only multigraphs appear as possible inputs to the reduction in the proof of \cref{lem:coloring-hard}, where parallel edges are removed explicitly. In the depth-three reduction, a bipartition $(S,T)$ separates source wires from target wires and suggests the gate orientation $S\to T$ in the intended circuit; candidate circuits are never assumed to respect this bipartition or orientation.

The \emph{square} $G^2$ of a graph $G=(V,E)$ has vertex set $V$, with two distinct vertices adjacent if their distance in $G$ is at most two. A \emph{tree decomposition} of $G$ consists of a tree $\mathcal T$ and bags $\{B_x\subseteq V\}_{x\in V(\mathcal T)}$ such that (i) every vertex of $G$ lies in some bag; (ii) every edge of $G$ has both endpoints in some bag; and (iii) for every $v\in V$, the nodes whose bags contain $v$ induce a connected subtree of $\mathcal T$ (the \emph{running-intersection property}). Its \emph{width} is $\max_x|B_x|-1$, and the \emph{treewidth} $\tw(G)$ is the minimum width of a tree decomposition of $G$.

\section{Recognizing and Constructing Two-Layer Circuits}
\label{sec:two}

We first show that clean ancillas give no advantage at depth two, which reduces the problem to the ancilla-free regime. We then characterize the combinatorial structure of two-layer circuits and use it to obtain an $O(n^2)$-time recognition and synthesis algorithm.

\begin{lemma}[Ancilla Invariance at Depth Two]
\label{lem:two-clean}
For every $M\in\GL(n,2)$ and every $m\ge0$, we have $d_m(M)\le2$ if and only if $d_0(M)\le2$.
\end{lemma}
\begin{proof}
The ``if'' direction holds by leaving all ancillas idle. Conversely, consider a two-layer circuit implementing $M$ on inputs $(x,0^m)$. An ancilla acting as a control in the first layer carries its initial value zero. A wire's value is unchanged when it serves as a control, so an ancilla acting as a control in the second layer must already be zero before that layer in order to finish at zero. Hence every ancilla-to-data gate is controlled by zero and has no effect on the data wires. Data-to-ancilla and ancilla-to-ancilla gates never modify data wires either, so deleting all gates incident to ancillas leaves two matching layers on the data wires that implement $M$.
\end{proof}

\subsection{Visibility of Gates and Bounded Treewidth of the Support Graph}

For an $n\times n$ matrix $M$, define its undirected \emph{off-diagonal support graph} $H_M=([n],E_M)$ by
\begin{equation}
 \{i,j\}\in E_M\ \Longleftrightarrow\ i\ne j\text{ and }(M_{ij}=1\text{ or }M_{ji}=1).
 \label{eq:support}
\end{equation}
This graph also appears as the vertex-connectivity graph in structural studies of CNOT synthesis~\cite{BuFanJoo2025}. Any two-layer ancilla-free implementation of $M$ has the form
\begin{equation}
 M=(I+Q)(I+P)=I+P+Q+QP,
 \label{eq:two-product}
\end{equation}
where $P,Q\in\F^{n\times n}$ are the off-diagonal matrices of the first and second matching layers ($P_{ab}=1$ if and only if the first layer contains $\cnot{b}{a}$, and similarly for $Q$).

\begin{lemma}[Gate Visibility]
\label{lem:visible}
If $M$ admits a two-layer ancilla-free implementation, then it admits one in which the endpoints of every gate form an edge of $H_M$.
\end{lemma}
\begin{proof}
Start with an arbitrary two-layer implementation of $M$. If the same directed gate $\cnot{j}{i}$ appears in both layers, the matching condition implies that neither $i$ nor $j$ takes part in any other gate, so the two gates cancel and can be removed without changing $M$.

After all such pairs are removed, consider a first-layer gate $\cnot{j}{i}$, so that $P_{ij}=1$. By vertex-disjointness, $P_{ij}$ is the only nonzero entry in column $j$ of $P$, so $(QP)_{ij}=\sum_k Q_{ik}P_{kj}=Q_{ii}P_{ij}=0$. Also, $Q_{ij}=0$ because identical gates have been removed. Entry $(i,j)$ of~\eqref{eq:two-product} gives $M_{ij}=P_{ij}+Q_{ij}+(QP)_{ij}=1$, so $\{i,j\}\in E_M$. Symmetrically, for a second-layer gate $\cnot{j}{i}$, $Q_{ij}=1$ is the only nonzero entry in row $i$ of $Q$, so $(QP)_{ij}=Q_{ij}P_{jj}=0$ and $P_{ij}=0$; hence $M_{ij}=1$.
\end{proof}

\begin{lemma}[Bounded Degree and Treewidth]
\label{lem:width}
If $d_0(M)\le2$, then $\Delta(H_M)\le4$ and $\tw(H_M)\le4$.
\end{lemma}
\begin{proof}
Fix a reduced two-layer implementation of $M$ as in \cref{lem:visible}, and let $J$ be the simple undirected graph formed by the edges of its two matchings. By \cref{lem:visible}, $J\subseteq H_M$. By~\eqref{eq:two-product}, an off-diagonal entry $M_{ij}=1$ arises from $P_{ij}=1$, from $Q_{ij}=1$, or from $(QP)_{ij}=1$, and the last case requires a vertex $k$ with $Q_{ik}=P_{kj}=1$. Every edge of $H_M$ therefore joins vertices at distance one or two in $J$, which gives
\begin{equation}
 J\subseteq H_M\subseteq J^2.
 \label{eq:sandwich}
\end{equation}
Since $J$ is a union of two matchings, its maximum degree is at most two, so each component is an isolated vertex, a path, or a cycle. Every vertex of $J^2$, and hence of $H_M$, has degree at most four.

It remains to bound the treewidth of $J^2$ componentwise. The square of a path $v_1,\dots,v_\ell$ has the path decomposition with bags $\{v_i,v_{i+1},v_{i+2}\}$, of width two. For a cycle $v_1,\ldots,v_\ell$ with $\ell\ge5$, the square $C_\ell^2$ has the path decomposition with bags
\[
 B_i=\{v_1,v_2,v_i,v_{i+1},v_{i+2}\},\qquad 3\le i\le\ell-2.
\]
The fixed vertices $v_1,v_2$ cover all wraparound edges of distance at most two, the sliding triples cover the remaining edges, and the running-intersection property holds; the width is four. Components with at most four vertices fit into a single bag. Hence $\tw(J^2)\le4$, and since treewidth is monotone under taking subgraphs, \eqref{eq:sandwich} gives $\tw(H_M)\le4$.
\end{proof}

The graph $H_M$ need not be a union of paths and cycles, nor equal to $J^2$. For instance, the circuit with layers
\[
 L_1 = \{\cnot{1}{2},\,\cnot{3}{4}\},\qquad L_2 = \{\cnot{2}{3}\}
\]
has gate graph $J$ equal to the path $1\text{--}2\text{--}3\text{--}4$, whereas $H_M$ has edges $\{1,2\},\{2,3\},\{3,4\},\{1,3\}$ but not the distance-two edge $\{2,4\}\in E(J^2)$: the gates at $2,3,4$ occur in the wrong order to propagate coordinate $2$ to wire $4$. Our algorithm computes a tree decomposition of $H_M$ directly and never needs to reconstruct $J$.

\subsection{Reduction to a Binary Constraint Satisfaction Problem}

To decide whether $d_0(M)\le2$, the algorithm constructs $H_M$ and rejects if $\Delta(H_M)>4$. Otherwise each wire has at most four neighbors in $H_M$, and by \cref{lem:visible} we may restrict attention to circuits whose gates lie in $E_M$. Then a wire $i$ has at most $9$ roles in a single layer: idle, control of one of its at most four neighbors, or target of one of them. Across both layers, the local state of $i$ takes at most $9\times9=81$ values.

We impose the following constraints on the constraint graph $H_M$:
\begin{enumerate}[label=(\roman*)]
 \item \emph{Matching reciprocity:} for every edge $\{i,j\}\in E_M$ and each layer $\tau\in\{1,2\}$, wire $i$ declares itself the control of $j$ in layer $\tau$ if and only if $j$ declares itself the target of $i$ in layer $\tau$. Since each local state selects at most one operation per layer, reciprocity guarantees that both layers are matchings.
 \item \emph{Target row equations:} for each wire $i$, let $a_i,b_i\in N_{H_M}(i)\cup\{\bot\}$ be the controls targeting $i$ in the first and second layers, as specified by the state of $i$ ($\bot$ meaning that $i$ is not a target). The $i$-th row of the product is then
 \begin{equation}
  e_i^\top
  +\indicator_{a_i\ne\bot}e_{a_i}^\top
  +\indicator_{b_i\ne\bot}e_{b_i}^\top
  +\indicator_{b_i\ne\bot,\ a_{b_i}\ne\bot}e_{a_{b_i}}^\top,
  \label{eq:row-constraint}
 \end{equation}
 which we require to equal $\row{M}{i}$.
\end{enumerate}

The row equation~\eqref{eq:row-constraint} needs constraints of arity at most two on $H_M$:
\begin{itemize}
 \item If $b_i=\bot$, the fourth term vanishes and the equation depends only on the state of $i$; it is a \emph{unary constraint}.
 \item If $b_i=j\in N_{H_M}(i)$, the index $a_{b_i}=a_j$ is encoded in the state of $j$, so the equation is a \emph{binary constraint} on the edge $\{i,j\}$.
\end{itemize}
Consistency between $j$ and its first-layer control $a_j$ is checked separately by the reciprocity constraint on $\{j,a_j\}$, so no row equation involves a third variable. The constraint graph is therefore $H_M$ itself rather than $H_M^2$.

\begin{lemma}[CSP Equivalence]
\label{lem:csp}
The constraint satisfaction problem above is satisfiable if and only if $M$ admits a two-layer ancilla-free CNOT implementation.
\end{lemma}
\begin{proof}
If $d_0(M)\le2$, \cref{lem:visible} provides a two-layer circuit supported on $H_M$, whose wire roles form a satisfying assignment. Conversely, given a satisfying assignment, reciprocity assembles the declared roles into two matching layers $I+P$ and $I+Q$, and the row constraints~\eqref{eq:row-constraint} ensure that every row of $(I+Q)(I+P)$ equals the corresponding row of $M$. No separate invertibility check is needed, since an accepted matrix is a product of matching layers.
\end{proof}

\subsection{Dynamic Programming and Running Time}

We run Bodlaender's linear-time algorithm~\cite{Bodlaender1996} on $H_M$ to test whether $\tw(H_M)\le4$ and, if so, to compute a tree decomposition of width at most four; if $\tw(H_M)>4$, we reject by \cref{lem:width}. With fixed domain size and fixed treewidth, the binary CSP can then be solved, with a witness, in linear time by standard dynamic programming~\cite{Freuder1990}. Every unary or binary constraint acts on a vertex or an edge of $H_M$, so it is contained in some bag; we assign each constraint to one such bag. Processing the rooted decomposition bottom-up, we tabulate for each bag which of the at most $81^5$ assignments to its at most five vertices are consistent with all constraints assigned to its subtree, using the standard introduce, forget, and join updates. The running-intersection property ensures that compatible local assignments combine into a global one, and backpointers recover a two-layer circuit whenever one exists.

For the running time, scanning the $n\times n$ matrix to build $H_M$ and the support of each row takes $O(n^2)$ time. If $\Delta(H_M)\le4$, then $H_M$ has at most $2n$ edges and every row of $M$ has weight at most five, so evaluating~\eqref{eq:row-constraint} for a fixed pair of states takes $O(1)$ time. Bodlaender's algorithm produces a decomposition with $O(n)$ bags in $O(n)$ time, and each bag is processed in time bounded by a constant depending only on the domain size $81$ and the bag size $5$. Together with \cref{lem:two-clean,lem:width,lem:csp}, this proves \cref{thm:two}.

\section{Structure of Three-Layer Clean-Ancilla Circuits}
\label{sec:normal}

Unlike at depth two, a three-layer circuit has just enough time to copy a data wire onto a clean ancilla in the first layer, use the ancilla as a nonzero control in the second layer, and erase it in the third. Our first lemma shows that this \emph{copy--use--clear} pattern is the only way ancillas can affect the data register: they add vertex-disjoint two-target fanouts to the middle layer, whose centers are idle in both outer data layers.

\begin{lemma}[Three-Layer Clean-Ancilla Normal Form]
\label{lem:normal}
For every $M\in\GL(n,2)$ and $m\ge0$, we have $d_m(M)\le3$ if and only if $M$ admits a factorization
\begin{equation}
 M=A_3(I+F)A_1,
 \label{eq:normal}
\end{equation}
where $A_1,A_3\in\GL(n,2)$ are matching layers on the $n$ data wires, and $F\in\F^{n\times n}$ is the adjacency matrix of a vertex-disjoint union of directed outward stars, each with one or two leaves. Moreover, the center of every two-leaf star in $F$ is isolated in both $A_1$ and $A_3$, and $F$ contains at most $m$ two-leaf stars.
\end{lemma}

Here $F_{ji}=1$ represents an effective middle-layer addition $\cnot{i}{j}$ from a star center $i$ to a leaf $j$. Thus $F$ has zero diagonal, every row of $F$ has weight at most one, every column has weight at most two, and no index has both a nonzero row and a nonzero column. In particular $F^2=0$, so $I+F$ is an involution, although a two-leaf star is not a matching layer on the data wires alone.

\begin{figure}[t]
\centering
\begin{tikzpicture}[x=1.15cm,y=.65cm,>=Latex]
 \foreach \y/\lab in {3/i,2/j,1/k,0/a_i}{
  \draw (0,\y)--(5,\y); \node[left] at (0,\y) {$\lab$};
 }
 \foreach \x in {1,4}{
  \draw (\x,3)--(\x,0); \fill (\x,3) circle (2pt);
  \draw (\x,0) circle (3.5pt); \draw (\x-.075,0)--(\x+.075,0); \draw (\x,-.13)--(\x,.13);
 }
 \draw (2.5,3)--(2.5,2); \fill (2.5,3) circle (2pt);
 \draw (2.5,2) circle (3.5pt); \draw (2.425,2)--(2.575,2); \draw (2.5,1.87)--(2.5,2.13);
 \draw (2.5,0)--(2.5,1); \fill (2.5,0) circle (2pt);
 \draw (2.5,1) circle (3.5pt); \draw (2.425,1)--(2.575,1); \draw (2.5,.87)--(2.5,1.13);
 \node[above] at (1,3.35) {copy}; \node[above] at (2.5,3.35) {use}; \node[above] at (4,3.35) {clear};
 \node[below] at (0,-.15) {$0$}; \node[below] at (5,-.15) {$0$};
\end{tikzpicture}
\caption{Realizing a two-leaf outward star $i \to \{j,k\}$ in the middle layer with one clean ancilla $a_i$. Wire $i$ is busy in layers one and three and therefore idle in $A_1$ and $A_3$.}
\label{fig:fanout}
\end{figure}
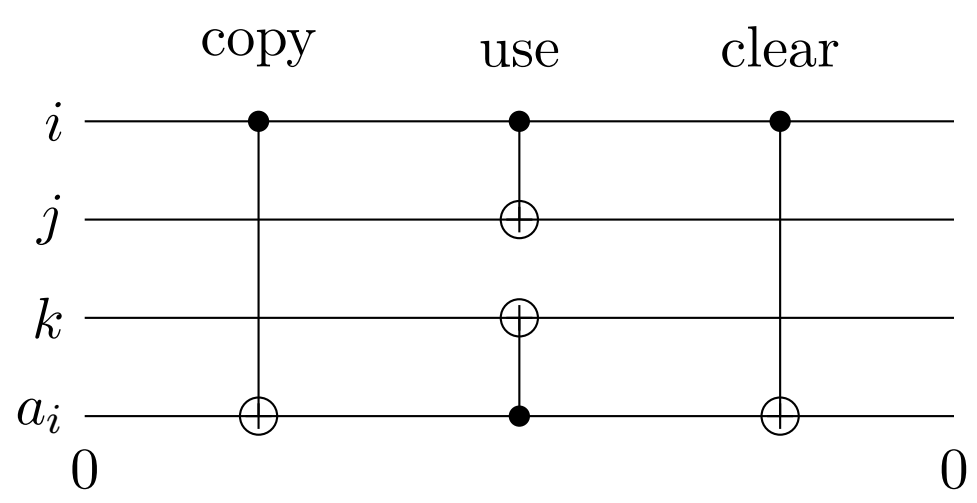

\begin{proof}
For necessity, pad a three-layer clean-ancilla circuit for $M$ to exactly three layers and restrict to inputs $(x,0^m)$, so that the state of each wire at any time is a linear form $w^\top x$ with $w^\top\in\F^{1\times n}$. Let $A_1$ be the matching layer formed by all first-layer data-to-data gates, with rows $u_1,\ldots,u_n\in\F^{1\times n}$; since $A_1\in\GL(n,2)$, these rows are linearly independent. Since all ancillas start at zero, a first-layer gate controlled by an ancilla acts trivially. Hence an ancilla is nonzero after layer one if and only if it is targeted by a copying gate $\cnot{i}{a}$ from a data wire $i$. Wire $i$ is then occupied in layer one and is idle in $A_1$, so its row after layer one is $u_i=e_i^\top$, and $a$ holds $u_i$. By the matching condition, each data wire is copied to at most one ancilla.

After the second layer, each data wire $j$ holds a row vector
\begin{equation}
 v_j=u_j\qquad\text{or}\qquad v_j=u_j+u_k\quad\text{for some }k\in[n],
 \label{eq:middle-row}
\end{equation}
where the second case occurs when $j$ is targeted by data wire $k$ or by an ancilla holding $u_k$ (the case $k=j$ gives $v_j=0$). Now consider an ancilla $a$ that acts as a nonzero control in the second layer, say holding $u_i$. Acting as a control leaves its state unchanged, so $a$ still holds $u_i$ after layer two. To return to zero, $a$ must be targeted in the third layer by a wire holding exactly $u_i$. This clearing control cannot be another ancilla $a'$, since $a'$ would then still hold $u_i\ne0$ at the output. Therefore $a$ is cleared by a data wire $j$ with $v_j=u_i$, and comparing with~\eqref{eq:middle-row} and using the linear independence of $u_1,\ldots,u_n$ forces $j=i$ and $v_i=u_i$.

This has two consequences for the source wire $i$ of an active ancilla $a$:
\begin{enumerate}[label=(\roman*)]
 \item Since $v_i=u_i\ne0$, wire $i$ is not targeted by any nonzero wire in the second layer (including its copy $a$).
 \item Wire $i$ is occupied by $\cnot{i}{a}$ in layers one and three, so it takes part in no data-to-data gate in either outer layer: $i$ is isolated in $A_1$ and in $A_3$, where $A_3$ is the matching layer of all third-layer data-to-data gates. (An ancilla acting as a control in layer three must already be zero to finish clean, so $A_3$ captures the entire third-layer action on the data register.)
\end{enumerate}

To construct $F$, keep all second-layer data-to-data gates, replace each second-layer gate $\cnot{a}{j}$ from a nonzero ancilla $a$ holding $u_i$ by the effective addition $\cnot{i}{j}$ (where $j\ne i$ by~(i)), and discard all gates that do not modify data wires. By the matching condition in layer two, each data wire is targeted at most once, so every vertex has in-degree at most one in $F$. Each data wire $i$ supplies at most one direct gate $\cnot{i}{j}$ and at most one ancilla-mediated gate (through its unique copy), so its out-degree is at most two. No vertex has both positive in-degree and positive out-degree: a direct control in layer two cannot also be a target in layer two, and an ancilla-mediated source is not targeted by any nonzero wire by~(i). Thus $F$ is a vertex-disjoint union of outward stars with one or two leaves. Each two-leaf star requires its own ancilla, so there are at most $m$ of them, and~(ii) shows that their centers are isolated in $A_1$ and $A_3$. This gives~\eqref{eq:normal}.

For sufficiency, given~\eqref{eq:normal}, assign a distinct clean ancilla $a_i$ to each two-leaf star $i\to\{j,k\}$ of $F$. As in \cref{fig:fanout}, execute $A_1$ together with $\cnot{i}{a_i}$ in layer one; $\cnot{i}{j}$, $\cnot{a_i}{k}$, and all one-leaf stars of $F$ in layer two; and $A_3$ together with the clearing gate $\cnot{i}{a_i}$ in layer three. Since two-leaf centers are isolated in $A_1$ and $A_3$ and the stars are vertex-disjoint, all three layers are matchings. Since $i$ is never targeted in the middle layer, it holds $x_i$ throughout and resets $a_i$ to zero in layer three; unused ancillas stay idle.
\end{proof}

\begin{corollary}[Workspace Saturation at Depth Three]
\label{cor:saturation}
For every $M\in\GL(n,2)$ and every $m \ge \lfloor n/3 \rfloor$, we have $d_m(M) \le 3$ if and only if $d_{\lfloor n/3 \rfloor}(M) \le 3$. In particular, \cref{thm:workspace} holds.
\end{corollary}
\begin{proof}
Each two-leaf star in $F$ occupies three data wires, so a vertex-disjoint collection on $n$ wires contains at most $\lfloor n/3\rfloor$ two-leaf stars. By the sufficiency part of \cref{lem:normal}, one clean ancilla per two-leaf star suffices, so $\lfloor n/3\rfloor$ ancillas suffice whenever $d_m(M)\le3$ for some $m$. The converse direction is trivial.
\end{proof}

More precisely, \cref{lem:normal} shows that the minimum number of clean ancillas needed to reach depth three equals the minimum number of two-leaf stars over all factorizations~\eqref{eq:normal} of $M$. This minimum can be positive.

\begin{example}[Clean ancillas help at depth three]
\label{ex:ancilla}
Consider the five-wire target
\[
 M=\begin{pmatrix}
 1&0&0&0&0\\
 1&1&0&0&1\\
 1&0&0&1&1\\
 1&0&1&1&0\\
 1&1&0&0&0
 \end{pmatrix}.
\]
With one clean ancilla $a$, the three layers
\[
 L_1=\{\cnot{1}{a},\,\cnot{5}{2},\,\cnot{3}{4}\},\qquad
 L_2=\{\cnot{1}{2},\,\cnot{5}{3},\,\cnot{a}{4}\},\qquad
 L_3=\{\cnot{1}{a},\,\cnot{2}{5},\,\cnot{4}{3}\}
\]
implement $M$ on the data wires and restore $a$ to zero; in the language of \cref{lem:normal}, wire $1$ is the center of the two-leaf star $1\to\{2,4\}$. Without ancillas, four layers suffice:
\[
 L'_1=\{\cnot{5}{2},\,\cnot{3}{4}\},\quad
 L'_2=\{\cnot{1}{2},\,\cnot{5}{3}\},\quad
 L'_3=\{\cnot{1}{4},\,\cnot{2}{5}\},\quad
 L'_4=\{\cnot{4}{3}\}.
\]
To show $d_0(M)>3$, we enumerated all matching layers on five wires (the identity, $20$ single-gate layers, and $60$ two-gate layers, $81$ in total) and computed the sets $\mathcal R_{k+1}=\{LX : X\in\mathcal R_k,\ L \text{ a matching layer}\}$ starting from $\mathcal R_0=\{I_5\}$. This gives $|\mathcal R_2|=4{,}211$ and $|\mathcal R_3|=123{,}781$, and $M\notin\mathcal R_3$; a Python script reproducing this enumeration is included as an ancillary file with the arXiv version of this paper. Since padding with identity layers makes $\mathcal R_k$ contain all products of at most $k$ layers, $d_0(M)=4$. By \cref{thm:two}, depth two is impossible even with ancillas, so $d_1(M)=3<d_0(M)=4$.
\end{example}

\subsection{Rigidity of High-Influence Fixed Sources}

Requiring $\row{M}{i}=e_i^\top$ does not by itself prevent a circuit from overwriting wire $i$ early and restoring it later. The next lemma shows that this cannot happen in the normal form if, in addition, $x_i$ influences at least four outputs of both $M$ and $M^{-1}$.

\begin{lemma}[Source Rigidity]
\label{lem:rigid}
Let $M\in\GL(n,2)$ and suppose that a wire $i\in[n]$ satisfies
\begin{equation}
 \row{M}{i}=e_i^\top,\qquad
 \wt(Me_i)\ge4,\qquad \wt(M^{-1}e_i)\ge4.
 \label{eq:rigid-hyp}
\end{equation}
Then in every factorization $M=A_3(I+F)A_1$ of the form~\eqref{eq:normal}, wire $i$ is not a target in $A_1$, $F$, or $A_3$; consequently, wire $i$ holds $x_i$ at every layer boundary.
\end{lemma}
\begin{proof}
Suppose that wire $i$ is targeted in $A_1$ by a gate $\cnot{j}{i}$ with $j\ne i$. After layer one, wire $i$ holds $e_i^\top+e_j^\top$, and it is still the only data wire whose state depends on $x_i$. Since $i$ is not isolated in $A_1$, \cref{lem:normal} forbids $i$ from being a two-leaf center, so $i$ has out-degree at most one in $F$.
\begin{itemize}
 \item If $i$ has out-degree zero in $F$, then $x_i$ appears on at most one wire after the middle layer, and the matching layer $A_3$ spreads it to at most two output wires, so $\wt(Me_i)\le2$.
 \item If $i$ has out-degree one in $F$, say via $\cnot{i}{k}$, then after the middle layer $x_i$ appears only on wires $i$ and $k$. Since $i$ has positive out-degree in $F$, it has in-degree zero, so wire $i$ still holds $e_i^\top+e_j^\top$. To produce $\row{M}{i}=e_i^\top$, wire $i$ must be targeted in $A_3$, so it cannot also act as a control there, while $k$ can pass $x_i$ to at most one further wire. Hence $x_i$ appears on at most three output wires, and $\wt(Me_i)\le3$.
\end{itemize}
Both cases contradict $\wt(Me_i)\ge4$, so $i$ is not targeted in $A_1$, and $\row{A_1}{i}=e_i^\top$.

Since $A_1$, $A_3$, and $I+F$ are involutions, inverting~\eqref{eq:normal} gives
\[
 M^{-1}=A_1(I+F)A_3,
\]
a factorization of $M^{-1}$ of the form~\eqref{eq:normal} with the roles of $A_1$ and $A_3$ exchanged. In addition, $\row{M}{i}=e_i^\top$ implies $\row{M^{-1}}{i}=e_i^\top$. The same argument applied to $M^{-1}$ with $\wt(M^{-1}e_i)\ge4$ shows that $i$ is not targeted in $A_3$, so $\row{A_3}{i}=e_i^\top$.

Finally,
\[
 e_i^\top = \row{M}{i} = \row{A_3(I+F)A_1}{i} = \row{(I+F)A_1}{i} = e_i^\top + \row{F}{i}A_1.
\]
Since $A_1$ is invertible, $\row{F}{i}=0$, so $i$ is not targeted in $F$ either.
\end{proof}

\section{Hardness at Depth Three via Edge Precoloring}
\label{sec:reduction}

\subsection{The Restricted Bipartite Edge-Precoloring Problem}
\label{sec:pins}

Our reduction starts from a restricted variant of bipartite three-edge-coloring with precolored leaves, denoted $\PinColor$. An instance consists of a finite simple bipartite graph $G=(S,T,E)$, a set $E_{\mathrm{pin}}\subseteq E$ of precolored edges, and a color assignment $c\colon E_{\mathrm{pin}}\to\{1,2,3\}$, subject to the following conditions:
\begin{enumerate}[label=(\roman*)]
 \item every source $p\in S$ has degree exactly three;
 \item every target $u\in T$ has degree one or three;
 \item every precolored edge $e\in E_{\mathrm{pin}}$, called a \emph{pin}, is incident to a degree-one vertex of $T$; and
 \item every source is incident to at most one pin, and if $p$ is incident to a pin, its other two neighbors have degree three.
\end{enumerate}
The question is whether $c$ extends to a proper three-edge-coloring of $E$ with colors $\{1,2,3\}$. Degree-three vertices of $T$ are called \emph{main targets}, and unpinned degree-one vertices of $T$ are called \emph{free leaves}. No girth restriction is imposed.

\begin{lemma}[Hardness of $\PinColor$]
\label{lem:coloring-hard}
$\PinColor$ is $\NP$-complete.
\end{lemma}

Three-edge-coloring is $\NP$-complete for cubic graphs~\cite{Holyer1981}, whereas bipartite graphs of maximum degree three are always three-edge-colorable~\cite{Konig1916}. With precolored edges, extension becomes $\NP$-complete for cubic bipartite graphs~\cite{Fiala2003}, even for planar ones~\cite{Marx2005}. These instances need not satisfy conditions~(iii) and~(iv), which our soundness analysis requires, but constant-size local gadgets convert them.

\begin{figure}[t]
\centering
\begin{tikzpicture}[x=0.95cm,y=0.75cm,thick,
    snode/.style={circle,draw,fill=black,inner sep=1.7pt},
    tnode/.style={rectangle,draw,fill=white,inner sep=2pt},
    lab/.style={font=\footnotesize}]
 \coordinate (y) at (0,1);
 \coordinate (h) at (1.5,1);
 \coordinate (a) at (2.8,2);
 \coordinate (b) at (2.8,0);
 \coordinate (h2) at (4.1,1);
 \coordinate (f) at (5.5,1);
 \coordinate (la) at (2.8,3.1);
 \coordinate (lb) at (2.8,-1.1);
 \coordinate (f1) at (6.5,1.8);
 \coordinate (f2) at (6.5,0.2);
 \draw (y)--node[above,lab] {$c$} (h);
 \draw (h)--(a)--(h2)--(b)--(h);
 \draw (a)--node[right,lab] {pin $c$} (la);
 \draw (b)--node[right,lab] {pin $c$} (lb);
 \draw (h2)--node[above,lab] {$c$} (f);
 \draw (f1)--(f)--(f2);
 \foreach \vertex in {y,a,b,f} \node[snode] at (\vertex) {};
 \foreach \vertex in {h,h2,la,lb,f1,f2} \node[tnode] at (\vertex) {};
 \node[left] at (y) {$y$};
 \node[above left] at (a) {$a$};
 \node[below left] at (b) {$b$};
 \node[below left] at (h) {$h$};
 \node[below right] at (h2) {$h'$};
 \node[below] at (f) {$f$};
\end{tikzpicture}
\caption{The forcing gadget $X_c(y)$ attached to a vertex $y\in S$. Circles denote vertices in $S$ and squares vertices in $T$. The two pinned leaf edges at $a$ and $b$ force the four cycle edges to alternate between the two colors other than $c$, so the edges $yh$ and $h'f$ both receive color $c$. The vertex $f$ has two free leaves.}
\label{fig:pin}
\end{figure}
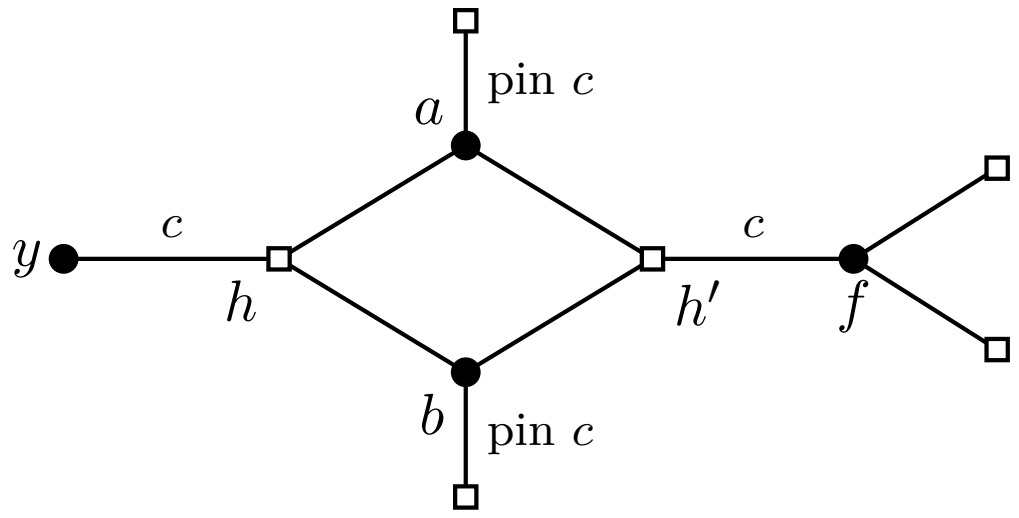

\begin{proof}
Membership in $\NP$ is clear. For hardness, we reduce from edge-precoloring extension on cubic bipartite graphs, which is $\NP$-complete by Fiala's theorem~\cite{Fiala2003}. Let $\Gamma=(S_0,T_0,E_0)$ be a cubic bipartite graph with a partial edge coloring $c_0$ defined on a set $E_{\mathrm{pre}}\subseteq E_0$.

\emph{Forcing gadget.} For a vertex $y\in S$ and a color $c$, the gadget $X_c(y)$ of \cref{fig:pin} consists of fresh vertices $a,b,f\in S$ and $h,h'\in T$, the four-cycle $h\,a\,h'\,b$, the edges $yh$ and $h'f$, a private pinned leaf of color $c$ at each of $a$ and $b$, and two free leaves at $f$. In a proper coloring, the edges $ah,ah'$ use the two colors other than $c$, and so do $bh,bh'$. Hence $ah$ and $bh$ receive distinct colors different from $c$, which forces $yh$ to have color $c$; likewise $h'f$ has color $c$. Conversely, if $yh$ is colored $c$, the coloring extends to the gadget by alternating the other two colors around the cycle, coloring $h'f$ with $c$, and coloring the free leaves at $f$ with the remaining two colors.

\emph{Replacing precolored edges.} For each $pq\in E_{\mathrm{pre}}$ with $p\in S_0$, $q\in T_0$, and $c_0(pq)=c$, let $\{c,c',c''\}=\{1,2,3\}$. Delete $pq$ and attach a gadget $X_c(p)$. Add a fresh vertex $z\in S$ adjacent to $q$, give $z$ a private pinned leaf of color $c'$, and attach a gadget $X_{c''}(z)$. At $z$, the pin uses $c'$ and the gadget edge is forced to $c''$, so $zq$ must have color $c$. Conversely, if $zq$ has color $c$, the coloring extends to $z$ and its gadget.

\emph{Removing parallel edges.} If parallel edges remain in $\Gamma$, keep one copy of each and replace every other copy $pq$ by a copy of $K_{3,3}$ minus an edge $x_0y_0$, with $x_0$ joined to $p$ and $y_0$ joined to $q$ (so $x_0$ lies in $T$ and $y_0$ in $S$). In a proper three-edge-coloring of this subgraph together with its two external edges, each color class matches equally many vertices on the two sides of $K_{3,3}$, so $x_0$ misses a color internally exactly when $y_0$ does; hence the two external edges receive the same color. Conversely, any common color $\gamma$ extends by three-edge-coloring the full $K_{3,3}$ with $\gamma$ on $x_0y_0$.

Let $G$ be the resulting graph, with pins consisting of the private pinned leaves. Every vertex of $S$ has degree three, every vertex of $T$ has degree one or three, and all pins are at leaves. The pinned vertices are the vertices $a,b$ of the gadgets, whose other neighbors $h,h'$ have degree three, and the vertices $z$, whose other neighbors are $q$ (of degree three again) and the vertex $h$ of $X_{c''}(z)$. Hence $G$ satisfies conditions~(i)--(iv), and its size is linear in that of $\Gamma$. If $c_0$ extends to $\Gamma$, coloring each new edge $ph$ and $zq$ with the color of the deleted edge $pq$ and extending into the gadgets gives an extension for $G$. Conversely, an extension for $G$ forces $ph$ and $zq$ to the color $c_0(pq)$; restoring each edge $pq$ with this color and contracting the parallel-edge gadgets yields a proper coloring of $\Gamma$ extending $c_0$.
\end{proof}

\subsection{Target Matrix Construction}
\label{sec:construction}

Given an instance $G=(S,T,E)$ of $\PinColor$, we construct a matrix $M_G\in\GL(N,2)$ on
\[
 N = |S| + |T| + 3|E_{\mathrm{pin}}|
\]
data wires. There is one wire for each vertex of $S\cup T$. For each pin $p\text{--}q$ (with $p\in S$, leaf $q\in T$, and color $c$), we add three private wires $r,s,t$, which together with $q$ form a four-wire timing gadget, or \emph{clock}, $(q,r,s,t)$. Distinct pins receive disjoint private wires.

Let $R$ be the set of all non-source wires (the unpinned targets in $T$ and the $4|E_{\mathrm{pin}}|$ clock wires), and order the coordinates with $S$ first and $R$ second. Then $M_G$ has the block lower-triangular form
\begin{equation}
 M_G=\begin{pmatrix}I&0\\K&H\end{pmatrix},
 \label{eq:MG}
\end{equation}
so every source wire keeps its input ($y_p=x_p$), and $K\in\F^{|R|\times|S|}$ and $H\in\GL(|R|,2)$ are defined as follows:
\begin{itemize}
 \item \emph{Unpinned targets:} for each unpinned target $u\in T$ (a main target or a free leaf), we prescribe
 \begin{equation}
  y_u=x_u+\sum_{p\in N_G(u)}x_p.
  \label{eq:ordinary-target}
 \end{equation}
 Thus $H$ acts as the identity on unpinned targets, and $\row{K}{u}$ is the indicator vector of $N_G(u)\subseteq S$.
 \item \emph{Clocks:} for each pin $p\text{--}q$ of color $c$ with clock $(q,r,s,t)$, define
 \begin{equation}
  H_0=\begin{pmatrix}
  1&0&0&0\\1&1&0&0\\1&0&1&0\\1&1&0&1
  \end{pmatrix},\qquad
  H_0^{-1}=\begin{pmatrix}
  1&0&0&0\\1&1&0&0\\1&0&1&0\\0&1&0&1
  \end{pmatrix},
  \label{eq:H0}
 \end{equation}
 and prescribe the four clock outputs
 \begin{equation}
  \begin{pmatrix}y_q\\y_r\\y_s\\y_t\end{pmatrix}
  =H_0\begin{pmatrix}x_q\\x_r\\x_s\\x_t\end{pmatrix}+w_c x_p,
  \qquad\text{where}\quad
  w_1=\begin{pmatrix}1\\1\\1\\1\end{pmatrix},\quad
  w_2=\begin{pmatrix}1\\0\\1\\0\end{pmatrix},\quad
  w_3=\begin{pmatrix}1\\0\\0\\0\end{pmatrix}.
  \label{eq:clock-output}
 \end{equation}
 No source other than $p$ contributes to the four rows of the clock.
\end{itemize}
Thus $H$ is the direct sum of the identity on unpinned targets and one copy of $H_0$ per clock; ordering each clock as $(q,r,s,t)$ makes $H$ and $M_G$ unit lower-triangular. The matrix $M_G$ depends only on $G$ and the pin colors, not on any coloring extension or ancilla budget, and it can be written down in $O(N^2)$ time.

\subsection{Completeness: From a Valid Coloring to a Three-Layer Circuit}

Suppose $G$ admits a proper three-edge-coloring extending the pins. For every unpinned edge $\{p,u\}$ of color $\tau$, schedule the gate $\cnot{p}{u}$ in layer $\tau$. For each pin $p\text{--}q$ of color $c$ with clock $(q,r,s,t)$, schedule the source injection $\cnot{p}{q}$ and the internal clock gates according to \cref{tab:clock}.

\begin{table}[t]
\centering
\begin{tabular}{cclll}
\toprule
Pin color $c$ & Source contribution & Layer 1 & Layer 2 & Layer 3\\
\midrule
1 & $w_1$ & $\cnot{p}{q}$ & $\cnot{q}{r}$ & $\cnot{q}{s},\ \cnot{r}{t}$\\
2 & $w_2$ & $\cnot{q}{r}$ & $\cnot{p}{q}$ & $\cnot{q}{s},\ \cnot{r}{t}$\\
3 & $w_3$ & $\cnot{q}{r}$ & $\cnot{q}{s},\ \cnot{r}{t}$ & $\cnot{p}{q}$\\
\bottomrule
\end{tabular}
\caption{Three-layer schedules for a clock $(q,r,s,t)$ attached to a pin $p\text{--}q$ of color $c$. The two internal stages implement $H_0$; injecting $\cnot{p}{q}$ before, between, or after them propagates $x_p$ with pattern $w_1$, $w_2$, or $w_3$.}
\label{tab:clock}
\end{table}

The two internal stages, $\cnot{q}{r}$ followed by $\{\cnot{q}{s},\cnot{r}{t}\}$, implement $H_0$ on $(x_q,x_r,x_s,x_t)^\top$. Injecting $x_p$ into $q$ before both stages, between them, or after both propagates it with column vector $H_0e_q=w_1$, $(I+E_{sq}+E_{tr})e_q=w_2$, or $e_q=w_3$, respectively, as required by~\eqref{eq:clock-output}. In \cref{tab:clock}, wire $q$ receives $\cnot{p}{q}$ in layer $c$ and takes part in internal clock gates only in the other two layers. Since the coloring is proper and clocks use disjoint private wires, no wire takes part in two gates of the same layer. Hence $d_0(M_G)\le3$, and $d_m(M_G)\le3$ for all $m\ge0$.

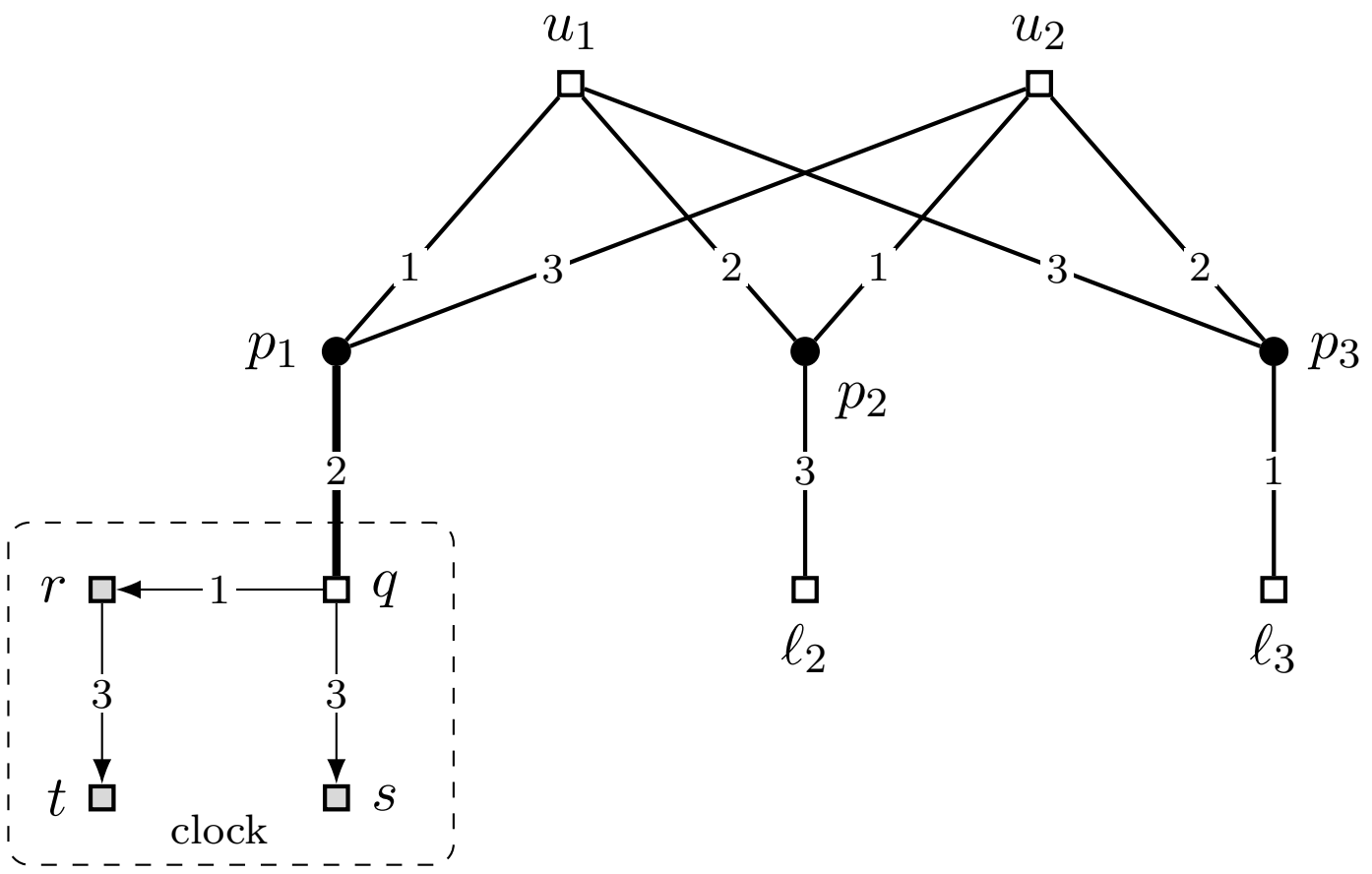
\begin{figure}[t]
\centering
\begin{tikzpicture}[x=1.05cm,y=1cm,thick,
    snode/.style={circle,draw,fill=black,inner sep=1.7pt},
    tnode/.style={rectangle,draw,fill=white,inner sep=2.2pt},
    cnode/.style={rectangle,draw,fill=black!15,inner sep=2.2pt},
    lab/.style={font=\scriptsize,fill=white,inner sep=1pt},
    >=Latex]
 \node[tnode,label=above:{$u_1$}] (u1) at (1.5,1.8) {};
 \node[tnode,label=above:{$u_2$}] (u2) at (4.5,1.8) {};
 \node[snode,label=left:{$p_1$}] (p1) at (0,0) {};
 \node[snode,label=below right:{$p_2$}] (p2) at (3,0) {};
 \node[snode,label=right:{$p_3$}] (p3) at (6,0) {};
 \node[tnode,label=right:{$q$}] (q) at (0,-1.6) {};
 \node[tnode,label=below:{$\ell_2$}] (l2) at (3,-1.6) {};
 \node[tnode,label=below:{$\ell_3$}] (l3) at (6,-1.6) {};
 \node[cnode,label=left:{$r$}] (r) at (-1.5,-1.6) {};
 \node[cnode,label=right:{$s$}] (s) at (0,-3) {};
 \node[cnode,label=left:{$t$}] (t) at (-1.5,-3) {};
 \draw (p1)--node[lab,pos=.3]{1}(u1);
 \draw (p1)--node[lab,pos=.3]{3}(u2);
 \draw[line width=1.6pt] (p1)--node[lab]{2}(q);
 \draw (p2)--node[lab,pos=.3]{2}(u1);
 \draw (p2)--node[lab,pos=.3]{1}(u2);
 \draw (p2)--node[lab]{3}(l2);
 \draw (p3)--node[lab,pos=.3]{3}(u1);
 \draw (p3)--node[lab,pos=.3]{2}(u2);
 \draw (p3)--node[lab]{1}(l3);
 \draw[->,thin] (q)--node[lab]{1}(r);
 \draw[->,thin] (q)--node[lab]{3}(s);
 \draw[->,thin] (r)--node[lab]{3}(t);
 \draw[dashed,thin,rounded corners] (-2.1,-1.15) rectangle (0.75,-3.45);
 \node[font=\scriptsize,anchor=south] at (-0.75,-3.45) {clock};
\end{tikzpicture}
\caption{The instance of \cref{ex:reduction}. Circles are sources and squares are targets; edge labels give a coloring extending the pin $p_1\text{--}q$ (bold, color $2$), and hence the layer of each source injection. The leaf $q$ is expanded into the clock $(q,r,s,t)$, whose internal gates (arrows, labeled by layer) follow \cref{tab:clock}; the clock wires $r,s,t$ are shaded.}
\label{fig:reduction}
\end{figure}

\begin{example}[A small instance]
\label{ex:reduction}
Let $G$ have sources $p_1,p_2,p_3$, main targets $u_1,u_2$ adjacent to all three sources, and one leaf per source: a pinned leaf $q$ at $p_1$ with color $2$, and free leaves $\ell_2,\ell_3$ at $p_2,p_3$ (\cref{fig:reduction}). This instance satisfies~(i)--(iv), and $M_G$ acts on $N=3+5+3=11$ wires. The main targets output $y_{u_j}=x_{u_j}+x_{p_1}+x_{p_2}+x_{p_3}$, the free leaves output $y_{\ell_i}=x_{\ell_i}+x_{p_i}$, and the clock outputs are
\[
 y_q=x_q+x_{p_1},\qquad y_r=x_r+x_q,\qquad y_s=x_s+x_q+x_{p_1},\qquad y_t=x_t+x_r+x_q,
\]
so $x_{p_1}$ appears exactly in outputs $q$ and $s$, matching $w_2=e_q+e_s$. The coloring of \cref{fig:reduction} yields the ancilla-free circuit
\[
 \begin{aligned}
 L_1&=\{\cnot{p_1}{u_1},\ \cnot{p_2}{u_2},\ \cnot{p_3}{\ell_3},\ \cnot{q}{r}\},\\
 L_2&=\{\cnot{p_1}{q},\ \cnot{p_2}{u_1},\ \cnot{p_3}{u_2}\},\\
 L_3&=\{\cnot{p_1}{u_2},\ \cnot{p_2}{\ell_2},\ \cnot{p_3}{u_1},\ \cnot{q}{s},\ \cnot{r}{t}\}.
 \end{aligned}
\]
A resynthesis pass might try to produce the pattern $w_2$ through other wires or ancillas; the soundness proof below shows that every three-layer implementation, with any number of clean ancillas, must place the injection $\cnot{p_1}{q}$ in layer two.
\end{example}

\subsection{Soundness I: Source Rigidity and Block Factorization}

Conversely, suppose that $d_m(M_G)\le3$ for some $m\ge0$, and fix a factorization $M_G=A_3(I+F)A_1$ as in \cref{lem:normal}. We first check that every source $p\in S$ satisfies the hypotheses of \cref{lem:rigid}. By~\eqref{eq:MG}, $\row{M_G}{p}=e_p^\top$, and
\[
 M_G^{-1}=\begin{pmatrix}I&0\\H^{-1}K&H^{-1}\end{pmatrix}.
\]
\begin{itemize}
 \item If $p$ is incident to no pin, it has three unpinned neighbors in $T$, on which $H$ and $H^{-1}$ act as the identity, so $\wt(M_Ge_p)=\wt(M_G^{-1}e_p)=4$.
 \item If $p$ is incident to a pin of color $c$, condition~(iv) ensures that its other two neighbors are main targets, contributing two ones to column $p$ of both $K$ and $H^{-1}K$. Its clock contributes $w_c\ne0$ to $K$ and $H_0^{-1}w_c\ne0$ to $H^{-1}K$. With the diagonal entry, $\wt(M_Ge_p)\ge4$ and $\wt(M_G^{-1}e_p)\ge4$.
\end{itemize}
By \cref{lem:rigid}, no source wire is targeted in $A_1$, $F$, or $A_3$. Hence, with respect to the partition $(S,R)$, the three factors are block lower-triangular:
\begin{equation}
 A_1=\begin{pmatrix}I&0\\U_1&V_1\end{pmatrix},\qquad
 I+F=\begin{pmatrix}I&0\\W&V_2\end{pmatrix},\qquad
 A_3=\begin{pmatrix}I&0\\U_3&V_3\end{pmatrix}.
 \label{eq:factor-blocks}
\end{equation}
Here $U_1,W,U_3\in\F^{|R|\times|S|}$ record source-to-$R$ additions in the three layers; since each wire of $R$ is targeted at most once per layer, every row of these matrices has weight at most one. The blocks $V_1,V_2,V_3\in\GL(|R|,2)$ describe operations within $R$; we do \emph{not} assume that they respect the boundaries between clocks and ordinary targets.

We say that $u\in R$ is \emph{isolated} in $V_\tau$ if $\row{V_\tau}{u}=e_u^\top$ and $V_\tau e_u=e_u$. The matching and disjoint-star constraints couple the blocks as follows:
\begin{enumerate}[label=(\alph*)]
 \item If $\row{U_1}{u}\ne0$, the matching condition on $A_1$ forbids $u$ from taking part in any gate within $R$ in layer one, so $u$ is isolated in $V_1$; in particular, $V_1U_1=U_1$.
 \item If $\row{W}{u}\ne0$, then $u$ is a leaf of a star in $F$. By the disjoint-star structure of $F$ (\cref{lem:normal}), $u$ has no other incoming edge and no outgoing edge in $F$, so $u$ is isolated in $V_2$.
 \item If $\row{U_3}{u}\ne0$, the matching condition on $A_3$ implies that $u$ is isolated in $V_3$.
\end{enumerate}

Multiplying out $M_G=A_3(I+F)A_1$ using~\eqref{eq:factor-blocks} gives $H=V_3V_2V_1$ and $K=V_3V_2U_1+V_3W+U_3$. By~(a), the first term equals $V_3V_2V_1U_1=HU_1$, and right-multiplying $H=V_3V_2V_1$ by the involution $V_1$ gives $HV_1=V_3V_2$. We use the three identities
\begin{equation}
 H=V_3V_2V_1,\qquad K=HU_1+V_3W+U_3,\qquad HV_1=V_3V_2
 \label{eq:block-identities}
\end{equation}
throughout the rest of the soundness proof.

\subsection{Soundness II: Main-Target Saturation and Source Disjointness}
\label{sec:targets}

We now analyze the rows of~\eqref{eq:block-identities} at main targets.

\begin{lemma}[Main-Target Saturation]
\label{lem:targets}
In every factorization~\eqref{eq:factor-blocks} of $M_G$, every main target $u\in T$ receives exactly one direct source injection in each of the three layers, one from each of its three neighbors in $N_G(u)$. Moreover, $u$ is isolated in $V_1$, $V_2$, and $V_3$.
\end{lemma}
\begin{proof}
Fix a main target $u$. By~\eqref{eq:ordinary-target},
\[
 \row{H}{u}=e_u^\top\qquad\text{and}\qquad \wt(\row{K}{u})=3.
\]
Substituting $\row{H}{u}=e_u^\top$ into the second identity of~\eqref{eq:block-identities} gives
\begin{equation}
 \row{K}{u} = \row{U_1}{u} + \row{V_3 W}{u} + \row{U_3}{u}.
 \label{eq:u-row-K}
\end{equation}
We first show that $u$ takes part in no gate within $R$ in the third layer.
\begin{itemize}
 \item Suppose $u$ is the control of a gate in $V_3$. Then $u$ is not targeted in $A_3$, so $\row{U_3}{u}=0$ and $\row{V_3}{u}=e_u^\top$, and~\eqref{eq:u-row-K} becomes $\row{K}{u}=\row{U_1}{u}+\row{W}{u}$, which has weight at most two. This contradicts $\wt(\row{K}{u})=3$.
 \item Suppose $u$ is targeted by a gate $\cnot{v}{u}$ in $V_3$ with $v\in R\setminus\{u\}$. Then $\row{U_3}{u}=0$ and $\row{V_3}{u}=e_u^\top+e_v^\top$, so
 \begin{equation}
  \row{K}{u}=\row{U_1}{u}+\row{W}{u}+\row{W}{v}.
  \label{eq:three-terms}
 \end{equation}
 Since $\wt(\row{K}{u})=3$ and each summand has weight at most one, all three summands are nonzero. By~(a), $u$ is isolated in $V_1$, and by~(b), both $u$ and $v$ are isolated in $V_2$. Evaluating row $u$ of $HV_1=V_3V_2$ gives
 \[
  \row{HV_1}{u} = e_u^\top V_1 = e_u^\top,
  \qquad\text{whereas}\qquad
  \row{V_3V_2}{u} = (e_u^\top + e_v^\top)V_2 = e_u^\top+e_v^\top,
 \]
 a contradiction.
\end{itemize}
Hence $u$ is isolated in $V_3$, and~\eqref{eq:u-row-K} becomes
\begin{equation}
 \row{K}{u}=\row{U_1}{u}+\row{W}{u}+\row{U_3}{u}.
 \label{eq:target-saturated}
\end{equation}
Since $\wt(\row{K}{u})=3$ and each term has weight at most one, the three terms have weight one and are supported on the three distinct neighbors of $u$. Since all three are nonzero, (a)--(c) show that $u$ is isolated in $V_1$, $V_2$, and $V_3$.
\end{proof}

The proof uses only the output row $\row{H}{u}=e_u^\top$ and the global identities~\eqref{eq:block-identities}. For example, if a circuit pre-combines two sources on an auxiliary wire $v\in R$ and forwards their sum to $u$ via $\cnot{v}{u}$, that gate also adds $x_v$ to $y_u$, which the identity $HV_1=V_3V_2$ rules out. Since main targets are isolated in $V_1,V_2,V_3$, they never forward source information to other wires of $R$.

\begin{lemma}[Source-Side Disjointness on Main Edges]
\label{lem:source-collision}
For every source $p\in S$ with at least two main-target neighbors, the direct injections from $p$ into its main-target neighbors occur in pairwise distinct layers. Moreover, $p$ is not the center of a two-leaf star in $F$.
\end{lemma}
\begin{proof}
In the normal form~\eqref{eq:normal}, a source $p$ can inject into two different wires of $R$ in the same layer only if $p$ is the center of a two-leaf star in $F$. Suppose this is the case. By \cref{lem:normal}, $p$ is isolated in $A_1$ and $A_3$, so the two star edges $\cnot{p}{v_1}$ and $\cnot{p}{v_2}$ are the only gates at $p$.
\begin{itemize}
 \item If $p$ has three main-target neighbors, \cref{lem:targets} requires $p$ to inject into all three, which is impossible with only two gates.
 \item If $p$ has exactly two main-target neighbors, they must be the leaves $v_1,v_2$. By \cref{lem:targets}, both are isolated in $V_3$ and cannot forward $x_p$ in layer three, so at the output $x_p$ appears only on $p$, $v_1$, and $v_2$. But the third neighbor of $p$ (a free leaf or a clock) must also receive a nonzero contribution from $x_p$ ($x_p$ itself or $w_cx_p\ne0$), a contradiction.
\end{itemize}
Hence $p$ is not a two-leaf center, and its injections into main targets occur in distinct layers.
\end{proof}

\subsection{Soundness III: Clock Rigidity}
\label{sec:clocks}

Now consider a pin $p\text{--}q$ of color $c$ with clock $(q,r,s,t)$. By condition~(iv), the other two neighbors of $p$ are main targets $u_1,u_2$. By \cref{lem:targets,lem:source-collision}, $p$ injects into $u_1$ and $u_2$ in two distinct layers and is not a two-leaf center. Hence $p$ has at most one remaining outgoing gate to $R$, in some layer $\tau\in\{1,2,3\}$.

Since $u_1,u_2$ are isolated in $V_1,V_2,V_3$ (\cref{lem:targets}), we have $He_{u_j}=V_3e_{u_j}=e_{u_j}$ for $j\in\{1,2\}$, so the injections into $u_1,u_2$ contribute exactly $e_{u_1}+e_{u_2}$ to column $p$ of $K=HU_1+V_3W+U_3$. Since $Ke_p=e_{u_1}+e_{u_2}+w_c$ (with $w_c\ne0$ supported on $(q,r,s,t)$), $p$ must execute one more gate $\cnot{p}{v}$ in layer $\tau$, targeting some $v\in R$ not assumed to lie in the clock, whose contribution on $R$ equals $w_c$:
\begin{equation}
 w_c = \begin{cases}
  H e_v & \text{if }\tau = 1,\\
  V_3 e_v & \text{if }\tau = 2,\\
  e_v & \text{if }\tau = 3.
 \end{cases}
 \label{eq:injection-times}
\end{equation}
(When $\tau=1$, the last two layers apply $V_3V_2$ to $e_v$, which equals $He_v$ because $V_1e_v=e_v$ by~(a).)

By~\eqref{eq:H0}, the four clock columns of $H$ are
\begin{equation}
 He_q=e_q+e_r+e_s+e_t=w_1,\qquad He_r=e_r+e_t,\qquad
 He_s=e_s,\qquad He_t=e_t,
 \label{eq:clock-columns}
\end{equation}
and every other column of $H$ is supported outside $\{q,r,s,t\}$. Hence $v=q$ is the only index in $R$ with $He_v=w_1$, no column of $H$ equals $w_2=e_q+e_s$ or $w_3=e_q$, and no column of $H$ has weight three. Every column of the matching matrix $V_3$ has weight at most two.

\begin{lemma}[Clock Rigidity]
\label{lem:clock}
For every pin $p\text{--}q$ of color $c$, the remaining source injection from $p$ is the gate $\cnot{p}{q}$, executed in layer $\tau=c$.
\end{lemma}
\begin{proof}
We use~\eqref{eq:injection-times} and~\eqref{eq:clock-columns}.

\paragraph{Case $c=1$.} Here $\wt(w_1)=4$. Since $\wt(V_3e_v)\le2$ and $\wt(e_v)=1$, we must have $\tau=1$ and $He_v=w_1$, which forces $v=q$.

\paragraph{Case $c=2$.} Here $w_2=e_q+e_s$ is neither a column of $H$ nor a unit vector, so $\tau=2$ and $V_3e_v=e_q+e_s$. Since $V_3$ is a matching, either $v=q$ (with $\cnot{q}{s}$ in $V_3$) or $v=s$ (with $\cnot{s}{q}$ in $V_3$). Suppose $v=s$. Since $s$ receives the middle-layer injection $\cnot{p}{s}$, property~(b) gives $V_2e_s=e_s$, and $HV_1=V_3V_2$ yields
\[
 H V_1 e_s = V_3 V_2 e_s = V_3 e_s = e_s + e_q.
\]
If $s$ is not a control in $V_1$, the left-hand side is $He_s=e_s$, a contradiction. Otherwise $V_1$ contains $\cnot{s}{z}$ for some $z\in R\setminus\{s\}$, and $He_s+He_z=e_s+e_q$ requires $He_z=e_q$, which is impossible. Hence $v=q$.

\paragraph{Case $c=3$.} Here $w_3=e_q$ is not a column of $H$, so $\tau\ne1$. If $\tau=2$, then $V_3e_v=e_q$ forces $v=q$ and $V_3e_q=e_q$. Since $q$ receives $\cnot{p}{q}$ in the middle layer, (b) gives $V_2e_q=e_q$, and $HV_1=V_3V_2$ yields
\[
 H V_1 e_q = V_3 V_2 e_q = e_q.
\]
If $q$ is not a control in $V_1$, this gives $He_q=e_q$, contradicting~\eqref{eq:clock-columns}. Otherwise $V_1e_q=e_q+e_z$ for some $z\in R\setminus\{q\}$, and substituting $He_q=e_q+e_r+e_s+e_t$ yields
\[
 H e_z = e_r + e_s + e_t,
\]
contradicting the fact that no column of $H$ has weight three. Hence $\tau=3$ and $e_v=e_q$, so $v=q$.
\end{proof}

\subsection{Decoding a Valid Three-Edge-Coloring}

We now extract from the factorization a proper three-edge-coloring of $G$ extending all pins:
\begin{enumerate}[label=(\roman*)]
 \item Each edge $\{p,u\}$ at a main target $u$ receives the layer in which $p$ injects into $u$ (\cref{lem:targets}).
 \item Each pin $p\text{--}q$ receives the layer of its injection $\cnot{p}{q}$, which equals its prescribed color by \cref{lem:clock}.
 \item For each unpinned source $p$ with free leaves, the free-leaf edges at $p$ receive the colors not used by the main-target edges at $p$.
\end{enumerate}

The coloring is proper. At every main target, the three incident edges receive distinct colors by \cref{lem:targets}; degree-one vertices impose no constraint. At a source $p$:
\begin{itemize}
 \item If $p$ is incident to a pin, its other two neighbors are main targets by~(iv), and, as shown at the start of \cref{sec:clocks}, its two main-target injections and $\cnot{p}{q}$ occur in three distinct layers.
 \item If $p$ is unpinned with at least two main-target neighbors, its main-target edges receive distinct colors by \cref{lem:source-collision}; with at most one main-target neighbor there is nothing to check. Step~(iii) then assigns the unused colors bijectively to the free-leaf edges.
\end{itemize}
This proves soundness for every clean-ancilla budget $m\ge0$ and completes the proof of~\eqref{eq:uniform-reduction}.

\section{Complexity Consequences}
\label{sec:complexity}

We combine the reduction of \cref{sec:reduction} with the structural results of \cref{sec:normal} to prove \cref{thm:three}, and then extend the result to variable depth thresholds.

\subsection{\texorpdfstring{$\NP$}{NP}-Completeness Across Clean-Ancilla Regimes}

\begin{proof}[Proof of \cref{thm:three}]
We first show membership in $\NP$. When the budget $m$ is polynomially bounded in $n$ (in particular, when $m=a(n)$ for a fixed polynomially bounded, polynomial-time computable function $a$), a witness for $d_m(M)\le3$ consists of three lists of CNOT gates on $n+m$ wires. A verifier checks that the gates in each list have pairwise disjoint supports, so that they form matching layers $L_1,L_2,L_3\in\GL(n+m,2)$, and evaluates the product on the clean-input subspace:
\[
 L_3 L_2 L_1 \begin{pmatrix}I_n\\0_{m\times n}\end{pmatrix} = \begin{pmatrix}M\\0_{m\times n}\end{pmatrix}.
\]
This takes polynomial time and verifies the clean-ancilla condition~\eqref{eq:clean} without constraining the blocks $B$ and $D$ in~\eqref{eq:extension}. In the unrestricted-ancilla model, \cref{cor:saturation} shows that $d_m(M)\le3$ for some finite $m$ if and only if it holds for $m=\lfloor n/3\rfloor$, so a witness needs at most $\lfloor n/3\rfloor$ ancillas and the same verifier applies.

For $\NP$-hardness, given an instance $G$ of $\PinColor$, the construction of \cref{sec:construction} produces in polynomial time a unit lower-triangular matrix $M_G\in\GL(N,2)$ satisfying~\eqref{eq:uniform-reduction} for every $m\ge0$. For a fixed budget function $a$, taking $m=a(N)$ shows that $d_{a(N)}(M_G)\le3$ if and only if $G$ is a yes-instance; likewise, $d_m(M_G)\le3$ for some finite $m$ if and only if $G$ is a yes-instance. By \cref{lem:coloring-hard}, both problems are $\NP$-hard, even for unit lower-triangular inputs. (To handle malformed inputs in a formal Karp reduction, map them to the fixed $9\times9$ unit lower-triangular no-instance $M_{\mathrm{no}}=I_9+\sum_{j=2}^9E_{j1}$. Each matching layer at most doubles the number of wires depending on $x_1$, so three layers reach at most $2^3=8$ wires even with ancillas, whereas $\wt(M_{\mathrm{no}}e_1)=9$; hence $d_m(M_{\mathrm{no}})\ge4$ for all $m\ge0$.)
\end{proof}

\subsection{Variable Depth Thresholds and Exact Optimization}

Hardness at the fixed threshold $k=3$ implies $\NP$-completeness when $k$ is part of the input, even in binary.

\begin{corollary}[Variable Depth Thresholds]
\label{cor:variable}
Deciding whether $d_0(M)\le k$ for an input pair $(M,k)$ with $k$ encoded in binary is $\NP$-complete. The same holds for inputs $(M,1^m,k)$ under variable polynomially bounded clean-ancilla budgets, and for $(M,k)$ under any fixed polynomially bounded, polynomial-time computable budget function $a(n)$.
\end{corollary}
\begin{proof}
$\NP$-hardness in all three settings follows from \cref{thm:three} with $k=3$. For membership, Gaussian elimination writes any $M\in\GL(n,2)$ as a product of at most $B(n)\coloneqq n(n-1)+3n$ CNOT gates without ancillas (at most $n(n-1)$ row additions and at most $n$ row swaps, each swap using three CNOTs). Hence $d_m(M)\le d_0(M)\le B(n)$ for all $M$ and $m$. If $k\ge B(n)$, the verifier accepts. Otherwise $k=O(n^2)$, each layer has at most $(n+m)/2$ gates, and a circuit of depth at most $k$ has polynomial size and can be verified as in the proof of \cref{thm:three}.
\end{proof}

Consequently, computing $d_m(M)$, or synthesizing a circuit of depth $d_m(M)$, is $\NP$-hard in each of these ancilla models. The three-versus-four gap also rules out polynomial-time $\alpha$-approximation of the minimum depth for each budget model in \cref{thm:three} and every fixed $1\le\alpha<4/3$, unless $\PP=\NP$: on yes-instances, an output within the same budget must have depth at most three since $3\alpha<4$, whereas on no-instances every legal circuit has depth at least four.

\section{Discussion and Open Problems}
\label{sec:discussion}

\paragraph{From depth two to depth three.}
Matrix sparsity alone does not explain the tractability of depth two: our hard instances $M_G$ are unit lower-triangular with constant row and column weights. The dichotomy instead comes from two structural differences. First, two matchings confine $H_M$ between a graph $J$ of maximum degree two and its square, which bounds the treewidth, whereas three layers leave enough room for four-wire clocks that fix the layer of each source injection against arbitrary resynthesis. Second, the cleanup constraint forces every ancilla control at depth two to be zero, while at depth three it allows exactly the copy--use--clear pattern, which yields the disjoint-star normal form (\cref{lem:normal}) and the $\lfloor n/3\rfloor$ workspace bound (\cref{cor:saturation}). For a fixed depth $k\ge4$, a causal-cone argument shows that at most $n(2^k-1)$ clean ancillas can ever become nonzero, since each matching layer at most doubles the number of wires depending on the data inputs and all other ancillas can be discarded. With four or more layers, however, ancillas admit multi-step parity transfers and deferred cleanup, and tight workspace bounds or a normal form at larger constant depths remain open.

\paragraph{Practical aspects of the depth-two algorithm.}
The linear-time part of our depth-two algorithm has a large constant, coming from Bodlaender's treewidth algorithm and up to $81^5$ states per bag. We do not claim that it competes with dedicated heuristic or SAT-based tools on general circuits~\cite{GoubaultEtAl2021,PehamEtAl2023,ShaikVanDePol2024,ShaikVanDePol2025,WebsterEtAl2026}; \cref{thm:two} shows that exact depth-two recognition needs no SAT or ILP solver. The state space can be pruned substantially in practice: components of $H_M$ can be solved independently, vertices of degree less than four have far fewer than $81$ states, and the reciprocity and row equations eliminate most state pairs before dynamic programming starts. If the target is given in a sparse representation that lists all nonzero entries of $M$, including diagonal entries, an input-sensitive $O(n)$-time recognition algorithm is a natural refinement.

\paragraph{Open problems.}
\begin{itemize}
 \item \emph{Fixed depths $k\ge4$:} our reduction proves hardness at threshold three but does not classify deciding $d_m(M)\le k$ for fixed $k\ge4$. Extending the rigidity and clock framework to larger thresholds remains open.
 \item \emph{Other ancilla and gate models:} our timing arguments rely on the vertex-disjointness of CNOT layers, which forbids simultaneous fanout from a shared control, and on clean ancillas starting at zero. Borrowed (dirty) ancillas, which start in an unknown state and must be restored regardless, change both the achievable shallow transformations and the available shortcuts. Several synthesis frameworks allow input or output qubit permutations~\cite{GoubaultEtAl2021,ShaikVanDePol2024,WebsterEtAl2026}; restricted connectivity such as linear nearest-neighbor or grid architectures introduces routing constraints~\cite{KutinMoultonSmithline2007,WuEtAl2023,GoubaultMartiel2025}; and wider circuits with mid-circuit measurement and classical feed-forward can reduce the CNOT depth of structured families such as CNOT ladders to a constant~\cite{TserkisEtAl2026}. The exact depth complexity in these models is open.
\end{itemize}

\section*{Declaration of AI Assistance}
OpenAI GPT-6.1-Sol assisted in developing the technical results and proofs in \cref{sec:two,sec:normal,sec:reduction}. GPT-6.1-Sol and Gemini-4-Argon assisted with drafting and revising the manuscript, generating the figures and tables, developing the leaf-pin reduction in \cref{sec:pins}, and verifying \cref{ex:ancilla}. The authors checked all claims, proofs, and citations and take full responsibility for the content of this work.

\begingroup
\raggedright
\bibliographystyle{plainnat}
\bibliography{references}
\endgroup
\end{document}